\documentclass[12pt,preprint]{aastex}

\begin{document}

\title{Fundamentalist physics: why Dark Energy is bad for Astronomy}
\author{Simon D.M. White}
\affil{Max Planck Institute for Astrophysics, Garching bei
M\"unchen, Germany}
\bigskip
\bigskip
\noindent{\bf{Astronomers carry out observations to explore the
diverse processes and objects which populate our Universe. High-energy
physicists carry out experiments to approach the Fundamental Theory
underlying space, time and matter. Dark Energy is a unique link
between them, reflecting deep aspects of the Fundamental Theory, yet
apparently accessible {\it only} through astronomical observation.
Large sections of the two communities have therefore converged in
support of astronomical projects to constrain Dark Energy. In this
essay I argue that this convergence can be damaging for astronomy. The
two communities have different methodologies and different scientific
cultures. By uncritically adopting the values of an alien system,
astronomers risk undermining the foundations of their own current
success and endangering the future vitality of their field.  Dark
Energy is undeniably an interesting problem to attack through
astronomical observation, but it is one of many and not necessarily
the one where significant progress is most likely to follow a major
investment of resources.}}
\bigskip

The pursuit of a deeper truth, of a fundamental theory which underlies
all others, is a powerful motivator in physics. So too are curiosity
and awe at the richness of Nature, at the connectedness which allows
disparate and seemingly unrelated processes to produce order, beauty
and diversity from apparent chaos. The first motivation is, perhaps,
most evident in high-energy physics, where a ``theory of everything''
has periodically appeared within reach, occupying many of the most
talented theoreticians. The second is evident in more
interdisciplinary, less ``fundamental'' fields, solid-state physics,
evolutionary biology, or astrophysics.  Fundamentalists prize the
depth of their research, seeing it as a means to abstract from the
complexity of the world a Truth which embodies the ultimate foundation
of the physics of particles and fields, thus, by extension, of all
physics, chemistry and biology. Generalists, on the other hand, prize
breadth and interdisciplinarity which promote the perception and
appreciation of the many truths underlying complex phenomena. In their
view, the fundamental theory of everything will contribute nothing to
our understanding of the origin and nature of life.

\bigskip
The discovery of Dark Energy, a near-uniform field which appears to
dominate the energy density of the current Universe and to drive its
accelerated expansion, has led astrophysicists and high-energy
physicists to make common cause. The apparent properties of the Dark
Energy, in particular the extremely low associated energy scale, are
entirely unexpected in the standard model of particle physics and
extensions such as supersymmetry. This suggests to many that Dark
Energy may somehow reflect the unification of gravity with the other
fundamental forces, and hence, paradoxically, physics at energies far
above those that can be probed directly with accelerators. At present
it seems that the properties of Dark Energy can be explored only
through astronomical observations, in particular through precise
measurements of the recent expansion history of the Universe and of
the growth of cosmic structure. Such measurements require observation
of large samples of complex astronomical objects such as galaxies,
galaxy clusters and supernovae. In consequence, astronomers interested
in supernovae and in cosmic structure formation have been working
intensively with high-energy theorists and astrophysical cosmologists
to design projects which might achieve the required precision. Such
collaborations bring risks as well as benefits because of different
motivational backgrounds and different methodologies in the two
communities. It is on this issue that I wish to focus in this essay.

\bigskip

\section*{HST and WMAP}

\bigskip
A useful illustration of the contrasts between the motivation and the
{\it modus operandi} of the two communities is provided by two current
satellite telescopes: the Hubble Space Telescope (HST) and the
Wilkinson Microwave Anisotropy Probe (WMAP).  These two projects have
very different scale and duration (in particular, HST is well over 10
times as expensive as WMAP) but this is not the aspect of the missions
which concerns me here. Both have been extremely successful and have
had very substantial impact both in the professional scientific
community and among the scientifically minded community at large. They
can serve as exemplars of supremely successful projects driven
primarily by traditional astrophysicists in the case of HST, and by
fundamental physicists and cosmologists in the case of WMAP.

\bigskip
HST, a NASA-led collaboration between the North American and European
astronomical communities, was planned over many years as the first
true observatory in space. It was anticipated that it would make great
advances because its location above the atmosphere allows observations
at higher resolution, at shorter wavelengths, and with less
contaminating sky emission than is possible with ground-based optical
telescopes. Studies made before launch assessed how these capabilities
might advance research frontiers as perceived at that time, but a
clear expectation was always that opening new windows of observational
parameter space would trigger new and unexpected discoveries in areas
where little was previously known. By and large this expectation has
been borne out. HST is now primarily known for discoveries which were
not part of its original science case.

\bigskip
WMAP was proposed by a small and tight-knit group of scientists in
response to a NASA call for suggestions for a mid-sized mission. All
the proposers had a strong track-record of instrumental development
for studies of the Cosmic Microwave Background (CMB), and several of
them had worked on NASA's previous CMB satellite COBE. Over the period
1989-1993, COBE had demonstrated the near-perfect black-body nature of
the CMB and had detected the weak temperature variations predicted 20
years earlier to remain as visible echoes of the early fluctuations
from which all present structure grew. (The PI's of the two relevant
instruments were awarded the 2006 Nobel Prize in Physics for these
discoveries.) Theoretical predictions for structure in the CMB are
robust within the standard paradigm. Detailed measurements should not
only test this paradigm, but should also give precise measurements of
the geometry of the Universe, of its contents in ordinary matter, dark
matter and dark energy, and of some aspects of the process which
created all structure, perhaps during a very early epoch of cosmic
inflation. The WMAP team designed their experiment specifically and
optimally (given available technology) to map fluctuations of the kind
predicted by theory and measured on large angular scales by COBE.
After several years of operation they have now successfully achieved
this goal.

\bigskip
One way to contrast the nature of these two projects is through the images
which have become emblematic of their success. In figure 1 I show two of the
best known images from HST, the picture of a region of the Eagle Nebula often
referred to as the ``Pillars of Creation" and the very deep exposure of an
apparently blank piece of sky known as the Hubble Deep Field. Both of these
pictures have had enormous public and professional impact, achieving iconic
status.  Both would be considered beautiful by many viewers.  To my mind their
beauty lies in the complexity of the structures and in the way they resonate
with visually familiar images but in a new and striking context. The Pillars
of Creation are reminiscent of backlit thunderclouds, evoking the reappearance
of sunlight after a storm, yet they depict the hidden birthplaces of stars.
The Hubble Deep Field shows us galaxy images almost like those in coffee-table
photographic atlases, yet these apparently neighboring systems in fact stretch
back through time nine tenths of the way to the Big Bang. In both cases we see
rich and complex systems whose structure and evolution are evoked rather than
characterised by the images. Quantitative analysis is possible, but it serves
to construct approximate phenomenological models rather than to measure
well-defined physical parameters

\bigskip
The best known images from WMAP are shown in figure 2. One is a rendition of
the CMB sky with all foregrounds removed and with a dramatic colour code to
emphasise the tiny temperature fluctuations detected by the satellite. This
has no resonance with the familiar world, appearing more like a mathematical
construction. To a very good approximation it is a random noise field and the
statistical properties of its several hundred thousand pixels are adequately
described by a six-parameter model. The second image compares the angular
power spectrum of this sky map with a prediction based on an {\it a priori}
physical model where all six parameters have been adjusted to values which are
fully compatible with those expected from independent, non-CMB data. The
agreement is a triumphant affirmation of the power of physics as a description
of our world. Although the present universe may be complex, the early universe
was simple, and we can calculate the statistical properties of its structure
from first principles. Fitting the observed data puts tight constraints on
``fundamental'' properties of the universe such as its overall geometry, its
contents in dark energy, dark matter and ordinary baryons, and the process
from which all structure originated. These properties affect the later growth
of galaxies and stars, but the CMB sky offers no insight into the complex
regularities which characterise these systems.

\bigskip
The following table gives another view of the contrasting properties of HST
and WMAP which illustrates some of the differences in scientific culture which
concern me in this essay.

\begin{tabular}{cc}

                HST                &                WMAP \\
& \\
          An observatory        &               An experiment\\
    Designed for general tasks      &     Designed for a specific task\\
    Serving a diverse community     &  Serving a single, coherent community\\
 Programme built through proposals   &     Programme set at design\\
      Many teams of all sizes       &   A single moderately large team\\
    Many results unanticipated      &        Main results `planned'\\
  Nourishes astrophysics skills   &  Nourishes data-processing/\\
& statistics skills \\
   Public support as a facility     &     Public impact through results\\
\end{tabular}

\bigskip
Most of these contrasts are self-explanatory, but the last one may
deserve more comment. In the wake of the Columbia disaster the NASA
administration decided that the planned shuttle mission to service the HST was
too risky and the telescope must therefore be allowed succumb to its natural
degradation in orbit and instrumentation.  This caused a tremendous outpouring
of support, not only from almost the entire astronomical community, but also
from the media, from the general public, and from the astronauts themselves.
Largely in response to this, the servicing mission is again on the NASA
roster. Although the impact of WMAP's results was enormous, it seems unlikely
there would have been such an emotional ground-swell of support had NASA
decided to discontinue its operations after four years.
 
\bigskip
This broad public affinity for astronomy reflects widespread interest
in deep questions of origin and fate which earlier civilizations
addressed through creation myths. Similar emotional undercurrents
explain the preponderance of `space' themes in popular science fiction
and the remarkable world-wide community of amateur astronomers. The
latter unites enthusiasts across generations, across skill levels,
across social strata, and across national and cultural
boundaries. Amateur astronomers build their own telescopes, use them
to do research of significant if not forefront interest, and maintain
a lively and high-quality magazine literature featuring substantive
reviews of new results from professional research. Astronomy resonates
with the popular imagination through its combination of complexity and
regularity, of the familiar and the strange, as well as through its
extraordinary and seemingly limitless range of subjects for study,
from the beginning of time to the birth of stars, from the
peculiarities of black holes to those of planets, from the origin of
the elements to that of spiral galaxies, from dark energy to the
preconditions for life. The fact that it is hard to imagine an
enthusiastic amateur community devoted to high-energy physics is
another indicator of the cultural differences between the two fields.

\section*{The two cultures}

\bigskip
Astrophysics and high-energy physics have a number of common
features. Neither has any direct application to everyday life, even if
their instrumental and computational needs sometimes lead to
significant technological spin-offs.  Both deal with phenomena
on scales which differ vastly from those of normal human
experience. Both require very expensive equipment. Despite this, the
research communities in the two fields differ notably in their
attitudes, in their motivations, in their {\it modi operandi}, and in
the value systems by which they judge their work.

\bigskip
Astrophysics aims to understand the structure and behaviour of
inherently complex systems and as a result is interdisciplinary and
synthetic in character. An intuitive feeling for the interplay between
phenomena from many areas of physics is needed, for example, to model
the formation of a galaxy. High-energy physics, in contrast, is
reductionist, aiming to break phenomena down into ever more
fundamental and more abstract entities, discarding along the way
complexities which may mask the underlying Truth.  Thus
astrophysicists tend to be generalists, prizing breadth of knowledge,
while high-energy physicists tend to be specialists, prizing the depth
to which they probe the underlying structure of matter. In
experimental work astrophysicists seek many truths associated with
many phenomena, and the best forefront research is characterised by
diversity and opportunism. In particle physics the quest for the
fundamental Truth has led to a focus on a much smaller number of
`important' questions (the origin of mass, the unification of quantum
mechanics and general relativity...) and to the organisation of 
industrial-strength teams to address them. New insights in
astrophysical research appeal on many levels, intellectual, emotional
and aesthetic, but they rarely display the quasi-mathematical rigour
of major advances in particle physics such as the understanding of
asymptotic freedom or of the Higgs mechanism. Astrophysicists are
universalists, democratic in perceiving interest in all aspects of the
cosmos, while high-energy physicists are fundamentalists, cleaving to
the pursuit of the single Truth.

\bigskip
Many of these differences can be traced to the fact that theory has
traditionally been tested against reality through controlled
experimentation in high-energy physics, but through observation in
astrophysics. The remoteness and scale of astronomical systems
preclude control of initial or boundary conditions, while long
timescales make evolution unobservable in most individual objects.
Astronomers are forced to work with ``snap-shots" of non-ideal,
strongly interacting and complex systems. This has produced a research
strategy quite unlike that in fields where experimentation is
possible.  When planning major new astronomy facilities, the principal
design drivers are usually:
\begin{enumerate}
\item to complement and extend previous facilities; 
\item to maximise the discovery potential; and 
\item to minimise the risk of scientific failure.  
\end{enumerate}

The emphasis is on enlarging capabilities by opening previously
unexplored regions of observational parameter space (in wavelength,
angular resolution, sensitivity...) rather than on targetting a
specific scientific issue. The science case, for HST, as for most
major observatories, was based on a wide range of problems from many
areas of astrophysics. The astronomical community has, nevertheless,
always considered HST's principal value to be the availability of most
of its observing time for programmes proposed after launch by
individual research groups. Most astronomers no longer remember the
original science justification for HST or most of the Key Programmes
implemented to address it.

\bigskip
To some extent, these considerations also apply to the design of major
facilities for high-energy physics, but even a global facility such as the
Large Hadron Collider is only able to address a relatively narrow range of
problems and to conduct a small number of experiments, each carried out by a
large, international team of physicists. These experiments are set up largely
according to traditional physics methodology:

\begin{enumerate}
\item identify the potential capabilities of new instrumentation; 
\item identify issues that these capabilities might address; 
\item refine the design \\
 a) to address the important issues optimally,\\ 
 b) to exclude confusing factors.  \\
\end{enumerate}
Team members specialise in optimising particular aspects of the
experiment (magnets, detectors, data analysis...) and may work for
decades before seeing data. Such long-term efforts require structured
and hierarchical management, and few physicists outside the teams are
able to work directly with the data. This contrasts with HST where
science is primarily carried out through programmes that last a couple
of years from proposal to completion and are independent both of the
instrument teams and of the science case which justified instrument
construction. The HST model offers young scientists a much wider range
of opportunities for scientific creativity and visibility than most
major accelerator experiments.

\section*{Dark Matter and Dark Energy}

Over the last two decades a standard paradigm has emerged for the
evolution of cosmic structure. One of its most striking aspects is the
assertion that the current universe is dominated by two unexpected and
apparently independent components, Dark Matter and Dark Energy. The
need for unseen matter to explain the dynamics of galaxy clusters was
first pointed out by Fritz Zwicky in the 1930's, but only the last 25
years have seen wide acceptance of the idea that cosmic structure
growth is driven by a gravitationally dominant population of some new
kind of weakly interacting particle. General acceptance of the idea
that the current expansion of the universe is accelerated by some form
of Dark Energy is even more recent, although the Cosmological Constant
was introduced by Einstein as part of his theory of gravitation
and is a viable explanation of current observations.

\bigskip
Both Dark Matter and Dark Energy are seen as fundamental by high-energy
physicists as well as by astrophysicists. All currently viable elementary
particle candidates for dark matter require an extension of the standard model
of particle physics with the lightest supersymmetric particle being, perhaps,
the current favorite. If this were confirmed, it would prove that the early
universe was sufficiently hot that supersymmetry was unbroken. Dark energy
seems to require an even more radical extension of current theories, perhaps a
unification of quantum mechanics and general relativity in some form of
superstring theory. The current evidence for both dark components is purely
astronomical, and it appears that {\it only} astronomical observations provide
a means to constrain properties of Dark Energy. Thus, the experimental testing
of the hottest idea in current high energy physics depends to an
unprecedented degree on astronomers, and the two communities have collaborated
substantially in planning major new initiatives to address the issue.

\bigskip
From an astronomical point of view, however, the Dark Matter and Dark Energy
problems differ qualitatively in their richness and in their interaction with
the rest of the field. Dark Matter drives the formation of galaxies and galaxy
clusters and influences all aspects of their structure.  Its distribution can
be mapped directly using gravitational lensing, and can be inferred indirectly
both from the dynamics of galaxies and intergalactic gas, and from the
structure of fluctuations in the microwave background radiation. The current
favorite candidate, the lightest supersymmetric partner of the known
particles, should produce annihilation radiation which could be imaged by
planned gamma-ray telescopes. Dark Matter may soon be observed directly by
underground ``telescopes" which are rapidly improving their ability to measure
the occasional collisions of Dark Matter particles with ordinary matter, and
it may be detectable in experiments at the Large Hadron Collider. Dark Matter
studies thus impact directly on most aspects of extragalactic astronomy and
astrophysical cosmology, as well as stimulating astroparticle experiments and
research programmes at accelerators.

\bigskip
In contrast, Dark Energy studies have little or no impact on other areas of
astrophysics and experimental high-energy physics. Models have been proposed
in which Dark Energy interacts with Dark Matter, resulting in observable
effects on structure formation, but in most models the two components are
effectively independent of each other. The effects of Dark Energy are then
manifest only in the overall expansion history of the universe and in the
linear growth rate of irregularities. If Einstein's theory of gravity holds,
one of these functions can be derived from the other and all astronomically
accessible information about Dark Energy is then contained in a single
observable function, the expansion rate as a function of cosmic time. Current
data are all consistent with the expansion history expected if Dark Energy
behaves like a Cosmological Constant. Estimates of the current value of the
relevant dimensionless parameter are in the range $w\sim -1\pm 0.1$, where
$w=-1$ at all times for a Cosmological Constant. For astronomers, this means
that the expansion history is already well enough measured that further
refinement will produce at most minor shifts in the inferred history of cosmic
structure formation.

\bigskip
Thus, while clarifying the nature of Dark Matter has all the hallmarks
of a typical ``astrophysicist's'' problem, interacting with many other
aspects of the field and accessible by many routes, clarifying the
nature of Dark Energy is a ``fundamental'' problem, apparently
accessible only by a route which has little impact on the rest of
astrophysics.

\bigskip
This has two consequences which are important for my argument.  Further
tightening of constraints on the cosmic expansion history and on the growth of
fluctuations will not improve our understanding of the formation and evolution
of stars, galaxies and larger structures. This is now limited primarily by
uncertainties in the many complex and interacting astrophysical processes
involved. Conversely, these uncertainties may affect our ability to place
tighter constraints on Dark Energy. For example, type Ia supernovae are
currently our best probe of the cosmic expansion history, and planned
programmes will increase the size and redshift range of well-observed samples
to the point where purely statistical errors are small. Unfortunately, the
progenitors of higher redshift supernovae formed and exploded in younger
galaxies than their lower redshift counterparts, and this could plausibly
cause small redshift-dependent shifts in the properties of the supernovae or
of their immediate environments. Undetected shifts of this kind could confuse
the search for the Dark Energy signal and limit the precision with which it
can be measured.  Similar systematic errors potentially afflict all other
proposed probes, since all are based on complex astrophysical objects. Thus
more precise constraints on Dark Energy will not help us understand the
evolution of the objects which populate our universe, but our ignorance in
this area could frustrate our attempts to constrain Dark Energy.

\section*{So why is Dark Energy bad for astronomy?}

I come now to the crux of my argument: how the current emphasis on
Dark Energy as a principal driver of astrophysical research can
undermine not only the methodological basis of astrophysics, but also
its attractiveness to its best practitioners, to the most talented of
next-generation scientists, and to the public at large.  In my view,
such negative consequences can result from importing the alien culture
of high-energy physics, especially in combination with an independent
trend towards ``Big Science" which is currently afflicting astronomy.

\bigskip
The dangers I see are of three kinds: inappropriate risk assessment in
the design of major programmes; investment of scarce resources in
programmes which do not enable new astrophysics or promote advances
over a broad front; promotion of a fundamentalist value system and a
managed work culture which will make astronomy unattractive to the
brightest, most creative and most ambitious young scientists. Let me
discuss these in turn.

\bigskip
The remarkable advances made recently through studies of the microwave
background have convinced many that astrophysical cosmology provides a
new window on fundamental physics. These advances were possible
because the observed structure takes the form of linear perturbations
of a simple state, an infinite uniform mixture of a small number of
components with well understood interactions. The evolution of this
system can be treated rigorously and precisely. In addition,
foreground effects are providentially weak and so cause only minor
complications.  Fundamentalist physicists, drawn to cosmology by this
success, often fail to appreciate the uniqueness of the
circumstances. An interesting comparison is helioseismology, the study
of the structure of the Sun based on sound waves propagating through
it. Here also the perturbations are linear and propagate in a medium
where the relevant physics is fully understood. Here also careful
measurement has produced extremely precise results for the properties
of a very large number of modes. Conclusions at the level of
confidence and precision reached by CMB studies are precluded,
however, by the complexity of the underlying system. For example, the
initial fraction of heavy elements required for the current standard
model of solar evolution to reproduce the structure inferred from
helioseismology is almost twice the fraction measured in the Sun's
atmosphere by analysis of its spectrum. 

\bigskip
Astrophysical routes to a better understanding of Dark Energy all
involve complex systems: supernovae to trace the cosmic expansion
history; galaxies to outline ripples in the large-scale matter
distribution; galaxies as background sources to trace gravitational
lensing by the foreground mass distribution; galaxy clusters as
markers of the growth of cosmic structure. Astrophysical experience
suggests that the ultimate precision reached by such programmes will
be set by systematic effects, for example, progenitor or environment
evolution for supernovae, nonlinear and non-determinate relations
between observables and theoretical quantities for galaxies and galaxy
clusters. By their nature such systematics cannot be accurately
assessed in advance, and indeed they often remain unrecognised until
the programme is complete.  Estimates of the final precision of Dark
Energy experiments are thus based primarily on purely statistical
considerations and should be considered optimistic estimates of the
"best possible" result. Dark Energy enthusiasts, emboldened by CMB
successes, often fail to appreciate these limitations, believing that
sophisticated statistical analysis will enable the best possible
result to be approached. This exposes the community to the danger of
designing and carrying out a very expensive experiment to measure many
thousand supernovae, or to image a very large area of sky, only to
find that the resulting measurement of $w$ is only a modest
improvement over previous work because of astrophysical
systematics. If the experiment is of limited use for other
astrophysical purposes, then the funds will, in effect, have been
wasted. A problem for which the astrophysicists will surely be blamed!

\bigskip
The potential problem here reflects the combination of inappropriate
risk assessment -- what is the chance that the complexities of real
galaxies, clusters or supernovae will frustrate attempts to measure
the cosmic expansion and fluctuation growth histories precisely? --
with an inappropriate design strategy -- planning an experiment like
WMAP rather than an observatory like HST.

\bigskip
This brings me to the second danger: the impoverishment of
astrophysics by too heavy an emphasis on Dark Energy when planning the
next generation of major facilities. As already discussed, astronomers
traditionally limit risk when designing new instruments by
concentrating on the expansion of technical capabilities in
sensitivity, wavelength coverage, spatial or spectral resolution. This
enables progress on a wide variety of problems, particularly since
operation in observatory mode allows new projects to be proposed as
they are seen to be interesting. Some Dark Energy projects conform to
this strategy. For example, wide-angle X-ray and millimeter surveys
will not only identify very large samples of distant galaxy clusters,
but will also image much of the sky to a sensitivity and resolution
which has not previously been achieved at these wavelengths. In
addition, these facilities will probably operate at least to a limited
extent in observatory mode. Other Dark Energy projects, for example
those searching for supernovae or looking to measure baryonic features
in the large-scale galaxy and mass distributions, will not extend
previous sensitivity, resolution or wavelength limits.  Rather they
achieve the required precision by observing much larger
areas of sky than has previously been possible.  Such surveys may not
enable significant progress in other areas of astrophysics. For
example, deep photometric imaging of 2 square degrees of the sky has
already been completed and provides data for hundreds of thousands of
faint galaxies.  Rather few studies of the formation and evolution of
galaxies would benefit from the 1000 times larger but otherwise
similar samples provided by Dark Energy surveys.  Since existing
instrumentation can match the capabilities of these surveys, there is
also little incentive to operate them in observatory mode.

\bigskip
The potential danger here is evident. The convergence between
astronomers and fundamental physicists produces a powerful lobby in
favour of Dark Energy experiments. In the natural competition between
proposed large projects this works to the disadvantage of more
traditional observatories at X-ray, radio, ultraviolet or infrared
wavelengths. These command strong support only from astronomers, and
so may be delayed, perhaps indefinitely, by financial constraints
resulting from implementation of ``higher priority" Dark Energy
experiments. Astronomers will spend their time, energy and resources
on experiments which have little impact on their main areas of
research, while sacrificing the facilities which have traditionally
driven creativity, innovation and the advance of knowledge in their
field.

\bigskip
This leads to the third, and in my view most serious danger. By accepting the
fundamentalist view that Dark Energy is so important that clarifying its
nature is the overiding problem for current astrophysics, astrophysicists
betray the underlying culture of their field and undermine its
attractiveness both to future generations of creative scientists and to the
public at large. This is exacerbated by other sociological trends within
astrophysics which I now digress briefly to discuss.

\bigskip
In figure 3 I show bibliographic statistics, compiled from NASA's
Astrophysics Data System (ADS), to illustrate changes in astrophysics
and space science over the last 30 years. In 1975 about 8500 different
authors published a total of about 8900 papers in the refereed
professional literature. By 2006 the number of authors had quadrupled
but the number of papers had only doubled. On the other hand, the mean
number of authors per paper also doubled, so that the number of papers
signed by a typical astronomer remained constant at about 2 per year.
The size of the astronomical community has thus increased dramatically
and a drop in the mean productivity of its members has been masked by
the tendency for more individuals to sign each paper. In 1975 over
40\% of all papers in the major journals had a single author and fewer
than 3\% had 6 or more authors. In 2006 only 9\% of papers had a
single author while almost 28\% had 6 or more authors. This trend
towards team-based projects is undoubtedly real, but it is accentuated
by the use of citations as a measure of performance, a practice which
may influence another strong trend visible in figure 3: the reference
lists of refereed astrophysics papers increased in length by a factor
of 3.4 on average between 1975 and 2006. Since the number of citations
to individuals for a given year is the product of the number of
papers, the mean length of their reference lists, and the mean number
of authors for the referenced papers, it was clearly much easier to
get cited in 2006 than in 1975 or even in 1995!  As an extreme
example, the fourth ranked astrophysicist by citations to papers
published over the last decade has never written a first-author paper
for a refereed journal and has gained almost all his citations through
his right to sign official papers by a large collaboration in which he
played a purely functional role. The increasing number of such survey
collaborations, usually put together to justify large time investments
on major facilities, means that more and more astrophysicists work in
directed, quasi-functional roles, and that fewer achieve visibility
through truly creative science.

\bigskip
The concentration on large long-term projects has long dominated
accelerator physics. Dark Energy projects will further accelerate this
trend in astrophysics. Only with very large surveys can one hope for a
percent level specification of the cosmic expansion and structural
growth histories. Achievement of these primary survey goals
will have little impact on astrophysics beyond the Dark Energy issue,
and most survey researchers will need to concentrate on functional
tasks to assure adequate data quality and timely completion of the
project. Contrast this with the traditional, opportunistic style of
the best astronomical research, where individuals or small groups
think up new ideas or build new instruments and apply them to
situations where the scientific return seems likely to be greatest. A
forward-looking observatory development programme can ensure that
there are always new problems to address and new opportunities to
extend the scientific frontier.  This is an attractive model for young
researchers. They can have a major scientific impact already as
graduate students and there is a clear path for them to establish
themselves rapidly as independent players in an international and
exciting field.  Such opportunities are rare in big survey science,
particularly in many Dark Energy projects.

\bigskip
This then is the third problem. If assembly of the very large surveys
needed to constrain Dark Energy comes to dominate astronomical
research, then the development of other new capabilities will be
slowed, and opportunities to carry out creative individual research in
most areas of astrophysics will be reduced. This will make our subject
less attractive to the best and most ambitious young scientists, who
will look to make their mark in other domains, biophysics or
nanotechnology perhaps.  Concentration on a single ``fundamental"
issue rather on the traditional diversity of issues will also make
astronomy less attractive to the general public, undermining taxpayer
support for the expensive facilities needed to pursue our science.
Listening to the siren call of the fundamentalists may lose us both
the creative brains and the instruments that are needed to remain
vibrant.  Dark Energy is the Pied Piper's pipe, luring astronomers
away from their home territory to follow high-energy physicists down
the path to professional extinction.

\bigskip

\section*{What is to be done?}

None of the negative consequences I have just outlined need
necessarily follow from our current situation. My intention in this
essay has been to draw attention to the dangers of uncritically
accepting that astronomers should spend much of their energy and
resources trying to clarify the nature of Dark Energy, just because it
is perceived as a fundamental (perhaps {\it the} fundamental) problem
by high-energy physicists. In my view a hard-nosed cost-benefit
analysis is needed, recognising both the inherent limitations of
observational astrophysics and the substantial cultural differences
between the astronomy and high-energy physics communities.

\bigskip
Dark Energy is a deep and interesting puzzle which can be probed by
astronomical observations alone, but it is one interesting puzzle
among many and it may be one of the least likely to be ``solved". We
do not know if astronomers can deliver measurements of the hoped-for
precision, but even if they do, it seems likely that high-energy
theorists will construct many Dark Energy models consistent with the
observed expansion and structure growth histories. Dark Energy will be
constrained, many possibilities will be excluded, but many others will
remain. Astronomers must be aware of this and must balance the needs
of Dark Energy projects against those of the core areas of our field.
New observatories promote exploration throughout astrophysics.  They
nourish the diversity and provide the opportunities for individual
creativity which underlie its current flourishing. We must not be
seduced by the fundamental nature of Dark Energy (and by the
availability of new funding sources) into sacrificing the foundation
of our subject's strength.

\bigskip
Here are some suggestions for accepting Dark Energy as a prime subject
for astronomical study while embracing neither the fundamentalist view
that it is the most important problem of our time, nor the industrial
work patterns engendered by ``Big Science" surveys of the kind
required to significantly tighten constraints on its properties.

\begin{enumerate}
\item Astrophysicists should recognise the cultural differences
between their own field and high-energy physics. They should be
willing to argue that astronomical discoveries -- that the Universe
expands, that the chemical elements were built in stars, that black
holes exist and can be far more massive than the Sun, that galaxies
continually change form, that other planets orbit other stars --
although qualitatively different, are no less significant for humanity
than the clarification of the underlying nature of forces and
particles.  They should resist the fundamentalist argument that
searching for the ultimate structure of space, time and matter is
deeper and more basic, and thus takes intellectual priority over other
ways of extending our knowledge of the physical world.

\item 
Large astronomical projects, even those for which Dark Energy issues
are a prime science driver, should continue to be designed to push
back the frontiers in many areas of astrophysics. Supernova surveys
should store enough information to explore the supernova mechanism and
the relation of supernovae to the stellar populations from which they
form, as well to trace out the expansion history of the
Universe. Galaxy redshift surveys should take sufficiently good
spectra for a sufficiently well-defined set of galaxies that galaxy
evolution can be studied, in addition to measuring the characteristic
scale of galaxy clustering for use as a standard measuring rod. This
is simply good astronomical practice, spreading the risk to compensate
for the fact that astronomers cannot ensure ``proper" laboratory
conditions for their experiments.

\item
Prioritisation of projects should be based not only on the case for
their prime science goal, but also on the extent to which they will
enable future advances in astrophysics as a whole. In the case of Dark
Energy surveys, this means recognising that refinement of the
principal quantities to be measured, the cosmic expansion and linear
fluctuation growth histories, is unlikely to impact significantly on
other areas of physical cosmology. Thus, the enabling aspect of such
surveys will come mainly from other science.

\item
Large projects require large teams and long time-scales. The negative
effects of this on young scientists' opportunities for creativity can
be drastic and must be mitigated by promoting a diverse set of science
goals for exploration by young team members. Both the Two Degree Field
Galaxy Redshift Survey (2dFGRS) and the Sloan Digital Sky Survey
(SDSS) were originally set up with a relatively narrow set of primary
science goals, but the teams involved were eventually able to address
a very broad range of problems with their survey data, and many of
these efforts were led by the younger scientists. In the case of SDSS,
the release of the full survey data through a powerful, publicly
accessible database has allowed astronomers across the world to carry
out their own SDSS projects, thereby enhancing the whole community's
opportunities for individual creativity.

\item
Credit for scientific contributions must be clearly assigned to those
responsible for the original insights and for the creative aspects of
the enabling work. Hard work alone brings little progress, and
appropriate recognition is a prime incentive attracting creative
scientists to our field.  Current assessment culture in astrophysics
is based mainly on total citations to papers signed by a scientist,
regardless of whether (s)he is sole author or author number 47 out of
165. This encourages inflated author and reference lists which dilute
the visibility of creative work over and above the dilution already
caused by the trend towards large teams. This could be off-set in
part by greater reliance on first author citations (in astrophysics
the first author is usually the person with primary scientific
responsibility for a paper) and on normalised citations (where an
author is credited with $1/N$ for a citation to a $N$-author
paper). This would remove the temptation to inflate author lists and
provide a fairer comparison of the overall impact of individual
astrophysicists.  Unless we recognise them properly, those capable of
original and creative contributions will prefer other fields.

\item 
Astrophysicists should motivate their activities in their own cultural
context, not in that of high-energy physics. This is particularly
important when interacting with students and young scientists. Dark
Energy is undeniably a fascinating puzzle, but it is a high-energy
physics puzzle. The creativity in {\it understanding} Dark Energy will
not come from planned astronomical surveys. They will provide more
precise measurements of quantities that are already well enough known
for astrophysics. Although reaching such precision is a major
challenge, it is a challenge that offers little opportunity for
scientific creativity unless one is primarily interested in the
processing of large datasets or the statistics of data
analysis. Bright and ambitious students will decide to become
astrophysicists only if they see an opportunity to make high-impact
contributions as individuals.  Within Dark Energy surveys, such
opportunities will come mainly from studies of astronomical
objects. In the rush to gain a funding edge by giving projects a Dark
Energy label, it is essential to avoid giving the impression that the
astronomical science is ``secondary", of less significance or interest
than improved measurements of the cosmic expansion and structure
growth histories. Indeed, it could turn out that Dark Energy is more
complex (or different) than most models suppose, and that critical
clues to its nature emerge from traditional astronomical exploration
of the phenomenology of structure, rather than from these precision
measurements.
\end{enumerate}

Astronomy often claims to be the oldest of the physical sciences and
it has a broader cultural and intellectual resonance with educated
society than any other branch of physics. For this reason many
university departments see astronomy as an ambassador for physics,
providing the non-scientific public with some understanding of the
scientific method and drawing students into physics from a wide
catchment area. The attraction lies in astronomy's diversity, in its
combination of a lack of direct application to human society with
insights into the development not only of our own world, but also of
the larger cosmos in which it is embedded. These strengths are
different from and complementary to those of fundamental physics. The
continued vitality of astrophysics does not depend on its ability to
constrain the Deep Truth underlying all reality, but rather on its
ability to retain our own and our public's fascination with the
many-facetted views it offers of the processes which shaped our
Universe and of the objects which populate it.

\section*{Acknowledgements}

This essay grew out of a talk given in summer 2006 in the director's
Blackboard Lunch at the Kavli Institute for Theoretical Physics.  I'd
like to thank the director, David Gross, for the invitation to give a
talk and for his spirited debate of the talk he got.\footnote{This
debate can be viewed ``live'' at
http://online.kitp.ucsb.edu/online/bblunch/white1/} I'd also like to
acknowledge the unique atmosphere of this institute, which strives
with remarkable success to promote cross-fertilisation between all
branches of physics, fundamental and otherwise.  Finally, I would like
to thank Alberto Accomazzi of the Smithsonian/NASA Astronomical Data
System for his help in compiling the publication statistics shown in
figure 3.

\begin{figure}
\epsscale{1.05}
\plotone{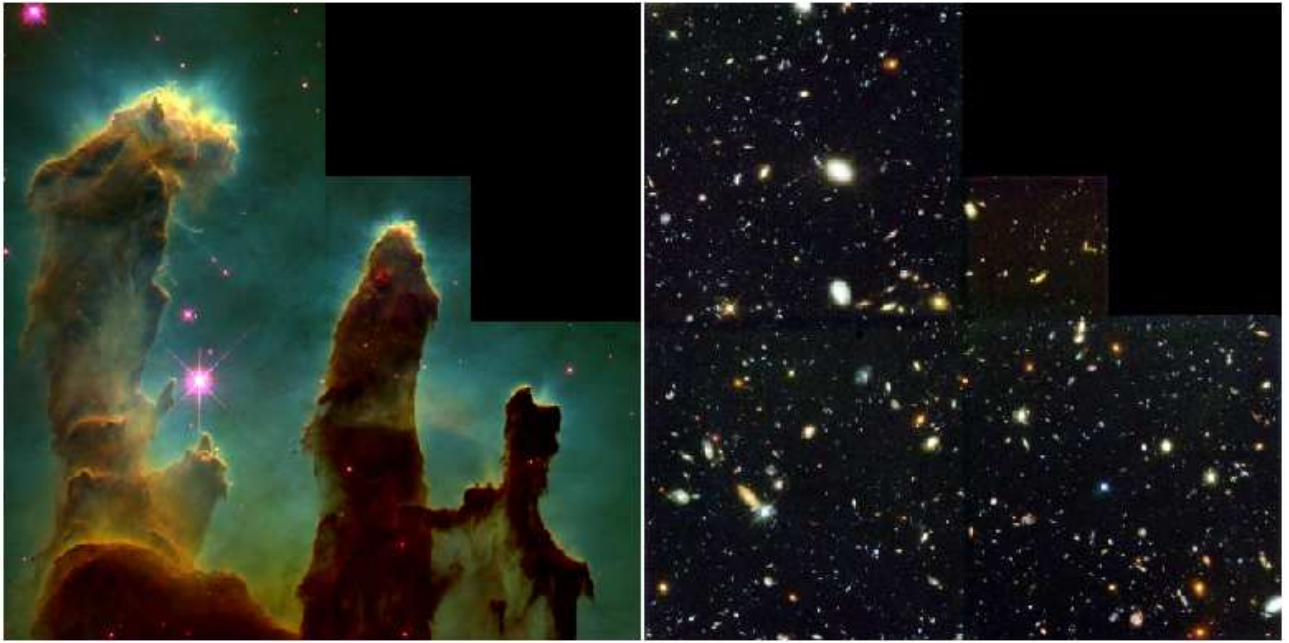}
\caption{Two emblematic pictures from the Hubble Space Telescope. On
the left is an image of the Eagle Nebula, a set of gas clouds
illuminated by young stars and enshrouding a number of stars in
formation. On the right is an image of the Hubble Deep 
Field.  At the time it was released in 1996 this was by
far the deepest image of the sky ever made, showing galaxies so 
distant that they are seen when the Universe was a small fraction 
of its present age.
\label{fig1}}
\end{figure}

\begin{figure}
\epsscale{0.8}
\plotone{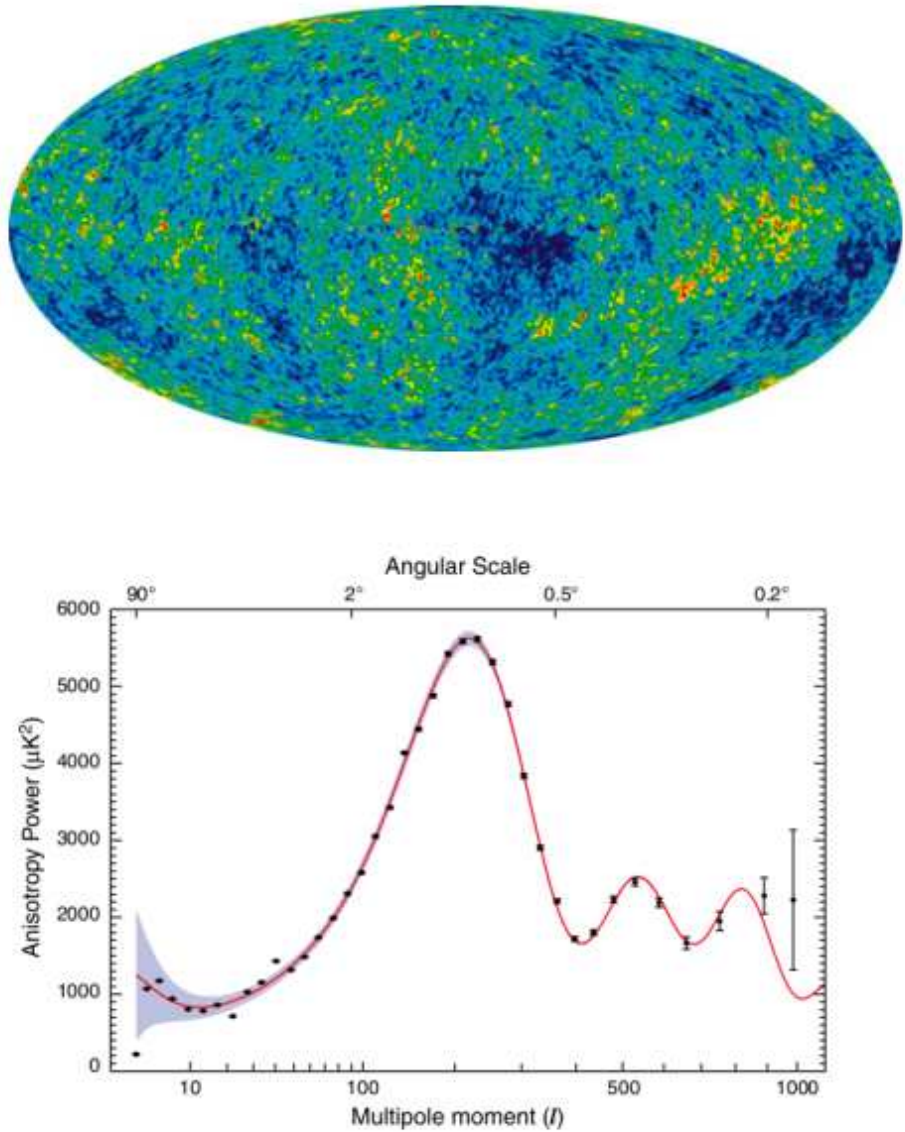}
\caption{ Two emblematic pictures from the Wilkinson Microwave
Anisotropy Probe. At the top is the WMAP map of temperature
fluctuations in the cosmic microwave background radiation. These
fluctuations are very weak, with typical amplitudes of a few parts in
100,000. They are a direct image of structure in the Universe when it
was only 400,000 years old. Below the map, its power spectrum
(the points with error bars) is compared with an {\it a priori} model
(the smooth curve) which assumes that all structure originated as
quantum zero-point fluctuations during a very early period of
inflationary expansion.  The six parameters specifying the model all
have physical meanings and they all take values which are quite
compatible with those inferred from independent astronomical
observations of the nearby Universe.
\label{fig2}}
\end{figure}

\begin{figure}
\epsscale{0.8}
\plotone{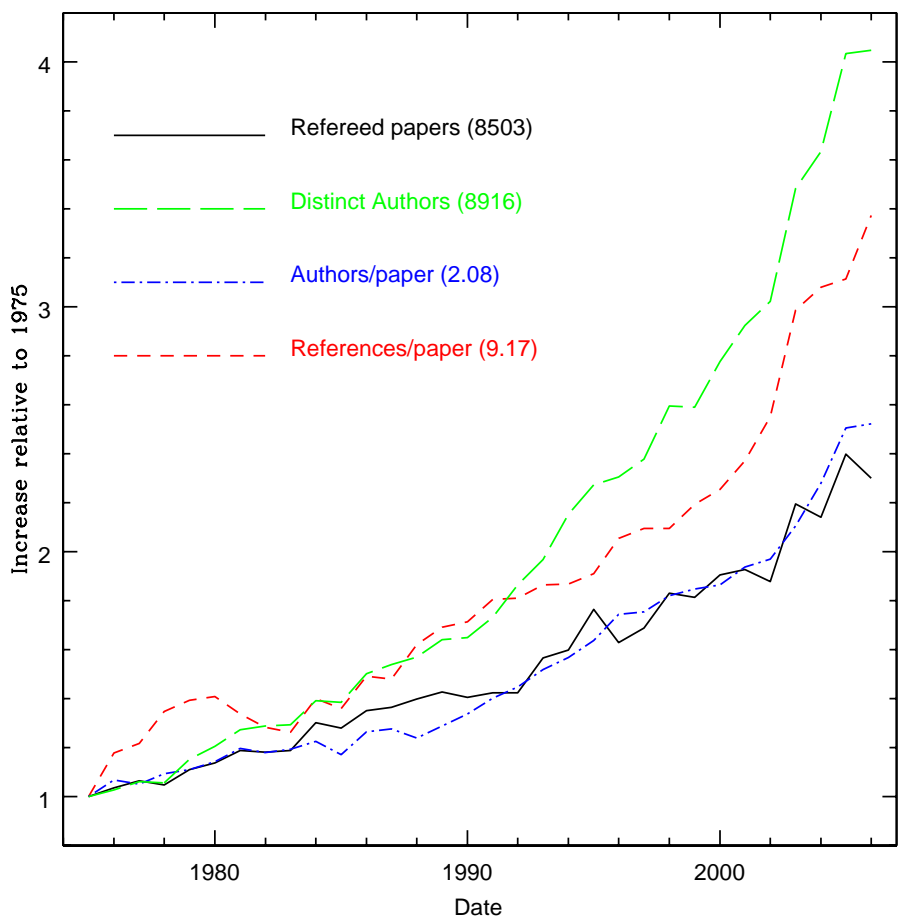}
\caption{Publication statistics based on data from the
Smithsonian/NASA Astrophysical Data System showing the growth of
activity in astrophysics over the period 1975-2006. The continuous
black line refers to the annual count of astrophysical papers in
refereed journals, the long-dashed green line to the total number of
distinct authors of these papers, the dot-dashed blue line to the
average number of authors signing a paper, and the short-dashed red
line to the average number of entries in each paper's reference
list. All statistics are normalised to be unity in 1975. The actual
1975 values for each are shown in parentheses after the labels in the
figure. Since 1975 the number of active astrophysicists has
quadrupled. The number of refereed papers and the number of authors
per paper has more than doubled.  The typical number of papers cited
has more than tripled.
\label{fig3}}
\end{figure}

\end{document}